\documentstyle[12pt]{article}

\makeatletter                    
\@addtoreset{equation}{section}  
\makeatother                     

\newskip\humongous \humongous=0pt plus 1000pt minus 1000pt
\def\caja{\mathsurround=0pt}
\def\eqalign#1{\,\vcenter{\openup2\jot \caja
        \ialign{\strut \hfil$\displaystyle{##}$&$
        \displaystyle{{}##}$\hfil\crcr#1\crcr}}\,}
\newif\ifdtup

\def\e{\eta}

\def\a{\alpha}
\def\b{\beta}
\begin{document}
\title{Backreaction in Semiclassical Cosmology: the Einstein-Langevin Equation}

\author{B. L. Hu\thanks{ Email: hu@umdhep.umd.edu}\\
{\small Department of Physics, University of Maryland,
College Park, MD 20742, USA} \\
A. Matacz\thanks{ Email: amatacz@physics.adelaide.edu.au}\\
{\small Department of Physics, University of Adelaide, 5005, Australia}}
\maketitle
\centerline{(umdpp 94-31) }

\begin{abstract}

Using the influence functional formalism we show how to derive a generalized
Einstein equation in the form of a Langevin equation for the description of
the backreaction of quantum fields and their fluctuations on the dynamics of
curved spacetimes. We show how a functional expansion on the influence
functional gives the cumulants of the stochastic source,
and how these cumulants enter in the equations of motion as noise sources.
We derive an expression for the influence functional in terms of the
Bogolubov coefficients governing the creation and annihilation operators
of the Fock spaces at different times, thus relating it to the difference in
particle creation in different histories. We then apply this to
the case of a free quantum scalar field in a spatially flat Friedmann-
Robertson-Walker universe and derive the Einstein-Langevin equations
for the scale factor for these semiclassical cosmologies.
This approach based on statistical field theory extends the conventional
theory of semiclassical gravity based on a semiclassical Einstein equation
with a source given by the average value of the energy momentum tensor,
thus making it possible to probe into the statistical properties of
quantum fields like noise, fluctuations, entropy, decoherence and dissipation.
Recognition of the stochastic nature of semiclassical gravity
is an essential step towards the investigation of the behavior of fluctuations,
instability and phase transition processes associated with the crossover
to quantum gravity.

\end{abstract}
\newpage
\section{Introduction}
Backreaction of quantum processes like particle creation in cosmological
spacetimes \cite{cpc} has been considered by many researchers
in the past for the
purpose of understanding how quantum effects affect the structure and
dynamics of the early universe near the Planck time \cite{scg,cpcbkr}.
Because of the general nature and complexity of the problem,
backreaction studies have also been used as a testing ground for the
development and application of different formalisms in quantum field theory
in curved spacetime \cite{BirDav}, e.g.,
regularization schemes to obtain finite energy-momemtum tensor,
perturbation methods, effective action formalism, etc. The most recent stage
of development for the discussion of cosmological backreaction problems
was the use of Schwinger-Keldysh (or closed-time-path, CTP) functional
formalism
\cite{ctp}, which, being formulated in the $in-in$ boundary condition,
gives rise to a real and causal equation of motion  ( the
semiclassical Einstein equation), where the expectation value of the
energy-momentum tensor of a quantum field acts as a source which drives
the classical effective geometry.  In this equation one can
identify a nonlocal kernel in the dissipative term whose integrated
dissipative power has been shown to be equal to the energy density of the
total number of particles created,
thus establishing the dissipative nature of the backreaction process
\cite{CalHu87,CalHu89}.

In pursuing a deeper understanding of the statistical mechanics meaning
of dissipation, one of us \cite{HuPhysica} cast this backreaction
problem into the conceptual framework of  quantum open systems \cite{qos}.
He made the observation that a Langevin-type equation is what should be
expected, and predicted that  for quantum
fields a colored noise source should appear in the driving term. He also
conjectured that the particle creation backreaction problem can be
understood succintly as the manifestation of a general fluctuation-dissipation
relation for quantum fields in dynamical spacetimes, a non-equilibrium
generalization of such relations depicting particle creation in black holes
\cite{CanSci,Sciama} and de Sitter universe \cite{Mottola}.
The missing piece in this search is the noise term associated with quantum
fields.

To look into this aspect of the backreaction problem in semiclassical gravity,
as well as exploring the quantum origin of noise and fluctuations
in inflationary cosmology \cite{HuBelgium}, and understanding the decoherence
problem in quantum to classical transition \cite{decrev},
Hu, Paz and Zhang \cite{HPZ1,HPZ2} looked into
the relation of colored noise and nonlocal dissipation in a quantum
Brownian motion model with the influence functional of Feynman and Vernon
\cite{FeyVer,CalLeg83,Gra}.
In this formalism the effects of noise and dissipation can be extracted from
the noise and dissipation kernels in the real and imaginary parts of the
influence functional, their interrelation residing in the
fluctuation-dissipation relation obtained as a simple functional relation.
If one views the quantum field as the environment
and spacetime as the system in the quantum open system paradigm, then the
statistical mechanical meaning of the backreaction problem in semiclassical
cosmology can be understood more clearly \cite{HuPhysica}.
In particular, one can identify
noise with the coarse-grained quantum fields \cite{HMLA,HM2},
derive the semiclassical Einstein equation as a Langevin equation
\cite{nfsg}, and understand the backreaction process as the
manifestation of a fluctuation-dissipation
relation \cite{fdrc}. Continuing their investigation of the backreaction
problem
via the CTP formalism, Calzetta and Hu \cite{nfsg} also found that the results
obtained from the influence functional formalism is the same as that
obtained earlier (but displayed only partially) from the
Schwinger-Keldysh method.  This paradigm has also been applied
to problems in quantum cosmology \cite{PazSin}.
(For an account of the search and discovery of
these ideas, see \cite{HPS,HuWaseda}.)

The specific goal of this paper is to derive the semiclassical Einstein
equation in the form of a Langevin equation.
Our primary task is the derivation of noise from the
quantum field source, and we do this by carrying out a cumulant expansion
of the influence functional.
This goal is shared by two other companion papers
addressing different aspects of this problem:
In Ref. \cite{nfsg}, using the closed-time-path method \cite{CalHu87,CalHu89},
Calzetta and Hu identified the source of decoherence and
particle creation to the noise kernel and showed their relation
through the Bogolubov coefficients. They also showed the relation of
quantum noise with classical fluctuations, and derived the semiclassical
Einstein equation with a noise term. In Ref. \cite{fdrc} Hu and Sinha
started with the density matrix of the universe
in quantum cosmology in the manner of \cite{PazSin} and demonstrated the
existence of a fluctuation-dissipation relation for the
particle creation and backreaction problem in a Bianchi Type-I universe.
These two papers together with this one present a quantum open system approach
to the backreaction problem in semiclassical gravity and cosmology. Together
they can
serve as a platform for exploring the transition to quantum cosmology.
It can also address the dissipative nature of effective theories
\cite{HuPhysica,eft},
and (to the extent that Einstein's general relativity can be viewed
as an effective theory) possible  dissipative effects in the low-energy
limit of any theory of quantum gravity. For a general discussion of these
ideas, see \cite{HuBanff}.

This paper is organized as follows:
In Sec. 2 we give a brief review of the influence functional formalism, mainly
to establish notations. Readers familiar with it can skip to the next section.
In Sec 3 we show how a functional expansion on the influence functional
gives the cumulants of the stochastic source, and how these cumulants enter
into the equations of motion as noise sources. In Sec. 4, following \cite{HM2},
we derive the form of the Hamiltonian for a scalar field in terms of its normal
modes and consider a class of actions where the field modes are coupled
parametrically to the scale factor of the universe. We derive an expression
for the influence functional in terms of the Bogolubov coefficients
governing the creation and annihilation operators of the Fock spaces at
different times which signify particle creation.
In Sec 5, we present two standard cases of
cosmological particle creation and derive the Einstein-Langevin equations
describing its backreaction on the background spacetime.

\section{Influence Functional Theory}
Consider the quantum system described by the action
$$
S[a,q]=S[a]+S_e[{\bf q}]+S_{int}[a,{\bf q}].
\eqno(2.1)
$$
We will consider $a$ to be our system variable and ${\bf q}$ to be our
environmental variables. Typically the environment has infinite
degrees of freedom which is denoted here by a bold type.

We will briefly review here the Feynman-Vernon
influence functional method for deriving the evolution operator.
The method provides an easy way to obtain a functional
representation for the evolution operator ${\cal J}_r$
for the reduced density matrix $ \hat\rho_r $.
Let us start first with the evolution operator ${\cal J}$ for the full density
matrix $ \hat\rho $ defined by
$$
\hat\rho(t)={\cal J}(t,t_i)\hat\rho(t_i). \eqno(2.2)
$$

As $\hat\rho$ evolves unitarily under the
action of (2.1), the evolution operator ${\cal J}$ has a simple path integral
representation. In the position basis, the matrix elements of the evolution
operator are given by

$$
\eqalign{
{\cal J}(a_f,{\bf q}_f,a'_f,{\bf q}'_f,t~
&|~a_i,{\bf q}_i,a'_i,{\bf q}'_i,t_i) =
{\cal K}(a_f,{\bf q}_f,t~|~a_i,{\bf q}_i, t_i)
  {\cal K}^*(a'_f,{\bf q}'_f,t~|~a'_i,{\bf q}'_i, t_i)\cr
&= \int\limits_{a_i}^{a_f}Da\int\limits_{{\bf q}_i}^{{\bf q}_f}D{\bf q}
{}~\exp{i\over\hbar}S[a,{\bf q}]
     \int\limits_{a'_i}^{{a'}_f} Da'\int\limits_{{\bf q}'_i}^{{\bf q}'_f}
D{\bf q}'~\exp{-{i\over\hbar}}S[a',{\bf q}']}                  \eqno(2.3)
$$
\noindent where the operator ${\cal K}$ is the evolution operator for the wave
functions. In a path-integral representation, the functional integrals are
over all histories
compatible with the boundary conditions. We have used
the subscripts $i, f$ to denote the initial and final variables at $t_i, t$.

The reduced density matrix is defined as
$$
\rho_r(a,a')
=\int\limits_{-\infty}^{+\infty}dq\int\limits_{-\infty}^{+\infty}
  dq'\rho(a,{\bf q};a',{\bf q}')\delta({\bf q}-{\bf q}')\eqno(2.4)
$$

\noindent and is propagated in time by the evolution operator ${\cal J}_r$
$$
\rho_r(a,a',t)
=\int\limits_{-\infty}^{+\infty}da_i\int\limits_{-\infty}^{+\infty}da'_i~
 {\cal J}_r(a,a',t~|~a_i,a'_i,t_i)~\rho_r(a_i,a'_i,t_i~).\eqno(2.5)
$$
By using the functional representation of the full density matrix evolution
operator given in (2.3), we can also represent ${\cal J}_r$ in path integral
form.
In general, the expression is very complicated since the evolution operator
${\cal J}_r$ depends on the initial state. If we assume that at a given
time $t=t_i$ the system and the environment are uncorrelated
$$
\hat\rho(t=t_i)=\hat\rho_s(t_i)\times\hat\rho_e(t_i),\eqno(2.6)
$$
then the evolution operator for the reduced density matrix does not depend
on the initial state of the system and can be written \cite{FeyVer} as
$$
{\cal J}_r(a_f,a'_f,t~|~a_i,a'_i,t_i)
 =\int\limits_{a_i}^{a_f}Da
   \int\limits_{a'_i}^{a'_f}Da'~
   \exp{i\over\hbar}\Bigl\{S[a]-S[a']\Bigr\}~{\cal F}[a,a'] \eqno(2.7)
$$
The factor ${\cal F}[a,a']$,  called the `influence functional', is defined as
$$
\eqalign{ {\cal F}[a,a']= &\int\limits_{-\infty}^{+\infty}d{\bf q}_f
 \int\limits_{-\infty}^{+\infty}d{\bf q}_i
 \int\limits_{-\infty}^{+\infty}d{\bf q}'_i
 \int\limits_{{\bf q}_i}^{{\bf q}_f}D{\bf q}
  \int\limits_{{\bf q}'_i}^{{\bf q}_f}D{\bf q}' \cr
& \times \exp{i\over\hbar}\Bigl\{
  S_e[{\bf q}]+S_{int}[a,{\bf q}]-S_e[{\bf q}']-S_{int}[a',{\bf q}'] \Bigr\}
  \rho_e({\bf q}_i,{\bf q}'_i,t_i) \cr
& = \exp{i\over\hbar} S_{IF}[a,a'] \cr}                \eqno(2.8)
$$
where $S_{IF}[a,a']$ is the influence action.
The effective action  for the open quantum system is defined as
$S_{eff}[a,a'] = S[a]-S[a'] + S_{IF}[a,a']$.

It is not difficult to show that (2.8) has the representation independent
form \cite{HM2}
$$
{\cal F}[a,a']=Tr\Bigl(\hat{U}[a_{t,t_i}]\hat{\rho}_e (t_i)
\hat{U}^{\dag}[a'_{t,t_i}]\Bigr) \eqno(2.9)
$$
where $\hat{U}$ is the quantum propagator for the action
$S_e[{\bf q}]+S_{int}[a(s),{\bf q}]$ where
$a(s)$ is treated as a time dependent classical forcing term.
We have found this form to be very convenient for deriving the influence
functional.

It is obvious from its definition that if the interaction term is zero, the
influence functional is equal to unity and the influence action is zero.
In general, the influence functional
is a highly non--local object. Not only does it depend on the time history,
but --and this is the more important property-- it also
irreducibly mixes the two sets
of histories in the path integral of (2.7). Note that the histories
$ a $ and $ a' $ could be interpreted as moving
forward and backward in time respectively.
Viewed in this way, one can see the similarity of the influence functional
\cite{FeyVer} and the generating functional in the closed-time-path
(CTP or Schwinger-Keldysh) integral formalism \cite{ctp}.
The Feynman rules derived in the CTP method are very useful for computing
the IF.

In those cases where the initial decoupling
condition $(2.6)$ is satisfied, the influence functional
depends only on the initial state of the environment. The influence functional
method can be extended to more general
conditions, such as thermal equilibrium between the system and the environment
\cite{HakAmb}, or correlated initial states \cite{CalLeg83,Gra}.

\section{Stochastic Forces from the Influence Functional}

In this paper we will be interested in models in which the action (2.1) has
a Hamiltonian of the general form
\begin{equation}
H(a,{\bf q})=H(a)+H_e({\bf q})+\sum_n\lambda \sigma (a,\dot{a})
\epsilon (q_n,\dot{q}_n)
\end{equation}
where $\lambda$ is a coupling constant and $\sigma$ and $\epsilon$ are
{\em arbitrary} functions of the system and environment variables.
The simplification made in (3.1) is that system environment interaction
is {\em separable}. This ensures that the effect of the environment on
the system can be described by a single stochastic source.

Let us introduce the sum and difference system variables as
\begin{equation}
\Sigma=\frac{1}{2}\biggl(\sigma(a,\dot{a})+\sigma(a',\dot{a}')\biggr),\;\;\;
\Delta=\sigma(a,\dot{a})-\sigma(a',\dot{a}'),
\end{equation}
 and define the real quantities
\begin{equation}
C_n(t_1,...,t_n;\Sigma_{t_1,t_i},...,\Sigma_{t_n,t_i}]=
\left(\frac{i}{\hbar}\right)^{-n}
\frac{\delta^n {\cal W}[\Sigma(s),\Delta(s)]}{\delta\Delta(t_1)...\delta
\Delta(t_n)}
\bigg|_{\Delta=0}
\end{equation}
where ${\cal W}={\rm ln}{\cal F}$.
The notation of $C_1(t_1;\Sigma_{t_1,t_i}]$ means $C_1$ is a function of
$t_1$ and also a functional of $\Sigma$ between endpoints $t_1$ and $t_i$.
Writing the influence action as a functional Taylor series and
generalizing the notation to $n$ variables we find that formally
\begin{eqnarray}
{\cal W}[\Sigma(s),\Delta(s)] & =
& \frac{i}{\hbar}\int_{t_i}^{t_f}dt_1\Delta(t_1)
C_1(t_1;\Sigma_{t_1,t_i}] \nonumber \\
&-&\frac{1}{2\hbar^2}\int_{t_i}^{t_f}dt_1\int_{t_i}
^{t_f}dt_2\Delta(t_1)\Delta(t_2)
C_2(t_1,t_2;\Sigma_{t_1,t_i},\Sigma_{t_2,t_i}]  \nonumber \\
& + & ...+\frac{1}{n!}\left(\frac{i}{\hbar}\right)^n\int_{t_i}^{t_f}dt_1...dt_n
\Delta(t_1)...\Delta(t_n)C_n(t_1,...,t_n;
\Sigma_{t_1,t_i},...,\Sigma_{t_n,t_i}] \nonumber \\
&+&...
\end{eqnarray}
This form of the influence functional will turn out to be useful for
its physical interpretation.
{}From (2.9) and the propagator $\hat{U}$ given by
\begin{equation}
\hat{U}[a_{t,t_i}]=\prod_n{\cal T}\exp\left[-\frac{i}{\hbar}\int_{t_i}^{t}ds
\Bigl(\hat{H}_{e}(\hat{q},s)
+\lambda \sigma(s)\hat{\epsilon}(\hat{q}_n,\dot{\hat{q}}_n)\Bigr)\right],
\end{equation}
it is observed that $C_n$ is of order $\lambda^n$ in the coupling constant.

We can interpret the $C_n$ in (3.4) as cumulants of a stochastic force.
Consider the action
\begin{equation}
S[a(s)]=\int_{t_i}^{t_f}ds\Bigl(L(a,\dot{a},s)
+\sigma(a,\dot{a})\xi(s)\Bigr)
\end{equation}
where $\xi(s)$ is some forcing term.
We want to consider the case where $\xi(s)$ is a stochastic force with
a normalised probability density functional ${\cal P}[\sigma(a,\dot{a});
\xi(s)]$. The probability
density functional is taken to be conditional on the system history
$\sigma(a,\dot{a})$.
The action (3.6) generates the influence
functional
\begin{eqnarray}
{\cal F}[\Sigma,\Delta]&=&\biggl\langle\exp\left[\frac{i}{\hbar}
\int_{t_i}^{t_f}
\xi(s)\Delta(s)ds\right]\biggr\rangle_{\xi} \nonumber \\
&\equiv& \int D\xi{\cal P}[\xi,\Sigma]\exp\left[\frac{i}{\hbar}
\int_{t_i}^{t_f}
\xi(s)\Delta(s)ds\right]
\end{eqnarray}
where $\Sigma$ and $\Delta$ are defined in (3.2).
The essential point is that the influence functional (3.7) has the exact
same form as the characteristic
function of the stochastic process $\xi$. Therefore given any influence
functional ${\cal F}[\Sigma,\Delta]$, we can interpret the $C_n$, given
by (3.3), as the cumulants of an effective stochastic force $\xi(s)$ coupled
to the system in a way described by the action (3.6). The probability
density functional ${\cal P}[\xi,\Sigma]$ of $\xi(s)$ can be obtained from
a given influence functional by inverting the functional fourier transform
(3.7).

If we ignore all cumulants after the second order
(the cumulants are of order $\lambda^n$) we are making a Gaussian
approximation to the noise. Although $\lambda$ is usually assumed to be small
for the series (3.4) to
converge, the formal expansion in orders of $\lambda$ is acceptable
even if $\lambda=1$, as long as the deviations from Gaussian are small.
With the Gaussian approximation we can write the influence functional as
\begin{eqnarray}
{\cal F}[a,a'] &=& \int D\xi{\cal P}[\xi,\Sigma] exp\left[{i\over \hbar}
S_{IF}[a,a',\xi]\right] \nonumber \\
&\equiv& {\left\langle \exp \left[{i\over \hbar}S_{IF}[a,a',\xi]\right]
\right\rangle}_{\xi}
\end{eqnarray}
where
\begin{equation}
{\cal P}[\xi,\Sigma] = P_0 \exp\Bigl(-\int\limits_{t_i}^{t_f}dt_1
\int\limits_{t_i}^{t_f}dt_2 ~\xi(t_1)
C_2^{-1}(t_1,t_2;\Sigma_{t_1,t_i},\Sigma_{t_2,t_i}]\xi(t_2) \Bigr)
\end{equation}
is the normalised functional distribution of $\xi(s)$ and
\begin{equation}
S_{IF}[a,a',\xi]=\int_{t_i}^{t_f}dt_1\Delta(t_1)
\Bigl(C_1(t_1,\Sigma_{t_1,t_i}]+\xi(t_1)\Bigr).
\end{equation}

We can use this effective action to obtain our semiclassical equation of
motion which is given by
\begin{equation}
\frac{\delta \Bigl(S_{eff}[a,a',\xi]\Bigr)}
{\delta \Delta_a (t)}\bigg|_{\Delta_a=0}=0
\end{equation}
where $\Delta_a=a-a'$. We find it becomes
\begin{equation}
\frac{\partial L}{\partial a}-\frac{d}{dt}\frac{\partial L}{\partial\dot{a}}
+\left(\frac{\partial \sigma}{\partial a}-
\frac{d}{dt}\frac{\partial \sigma}{\partial\dot{a}}\right)
\Bigl(C_1(t,\sigma_{t,t_i}]+\xi(t)\Bigr)-
\frac{\partial \sigma}{\partial \dot{a}}
\Bigl(\dot{C}_1(t,\sigma_{t,t_i}]+\dot{\xi}(t)\Bigr)=0
\end{equation}
where $L(a,\dot{a})$ is the system Lagrangian and
$\xi(t)$ is a zero- mean Gaussian stochastic force with a correlator given
by
\begin{equation}
\langle \xi(t)\xi(t')\rangle=C_2(t,t';\sigma_{t,t_i},\sigma_{t',t_i}].
\end{equation}

Clearly both the noise and driving term are still dependent on the system
history in a complex way in general. However we can further simplify things by
expanding around a background $a=a_b$. In this case we approximate
the first cumulant by
\begin{equation}
C_1(t;\sigma_{t,t_i}]=C_1(t;\sigma_{t,t_i}]|_{\sigma=\sigma_b}+
\int_{t_i}^{t} dt' \tilde\sigma(t')\mu(t,t')+...
\end{equation}
\begin{equation}
\dot{C}_1(t;\sigma_{t,t_i}]=\dot{C}_1(t;\sigma_{t,t_i}]|_{\sigma=\sigma_b}+
\int_{t_i}^{t} dt' \tilde\sigma(t')\dot{\mu}(t,t')+...
\end{equation}
where $\tilde\sigma=\sigma-\sigma_b$ and
\begin{equation}
\mu(t,t')=\frac{\delta C_1(t;\sigma_{t,t_i}]}
{\delta \tilde\sigma(t')}\bigg|_{\sigma=\sigma_b}
\end{equation}
\begin{equation}
\dot{\mu}(t,t')=\frac{\delta \dot{C}_1(t;\sigma_{t,t_i}]}
{\delta \tilde\sigma(t')}\bigg|_{\sigma=\sigma_b}
\end{equation}
where we have assumed in (3.15) that $\mu(t,t')$ is antisymmetric as will be
the case for our examples.
The noise $\xi(t)$ now has the correlator
\begin{equation}
\langle \xi(t)\xi(t')\rangle=
C_2(t,t';\sigma_{t,t_i},\sigma_{t',t_i}]|_{\sigma=\sigma_b}.
\end{equation}
These approximations greatly simplify (3.12).
Our task is then to solve for the fluctuations $\tilde{a} \equiv a- a_b$
subject to the initial condition $\tilde{a}(t_i)=\dot{\tilde{a}}(t_i)=0$.

\section{Influence Functional for Cosmological Backreaction}

In this section, following the methods of \cite{HM2}, we will derive the
form of the influence functional
in terms of the Bogolubov coefficients in the transformation between the
creation
and annihilation operators of field amplitudes at different times.
First we show how the dynamics of a general real scalar field in an expanding
FRW universe can be described by a sum over quadratic time dependent
Hamiltonians.
Then we discuss the Bogolubov coefficients in terms of the squeeze parameters
\cite{sqpc}.
It also applies to the case of gravity wave perturbations whose two
polarizations
obey wave equations of the same form as a massless, minimally coupled scalar
field (see \cite{GriSid} for details).

The action for a free scalar field in an arbitrary space-time can be written as
the sum of gravitation action $S_g$ and matter action $S_m$ of the form
\begin{equation}
S_g=l_p^2\int d^4x\sqrt{-g}(R-2\Lambda)-
2l_p^2\int d^3x \sqrt{-h}K
\end{equation}
\begin{equation}
S_m= \frac{l_p^2}{2}\int d^4x\sqrt{-g}
\left(g^{\mu\nu}\bigtriangledown_{\mu}\Phi\bigtriangledown_{\nu}\Phi-
(m^2+\xi_d R)\Phi^2\right)
+\xi_d l_p^2\int d^3x\sqrt{-h} K\Phi^2 .
\end{equation}
where $l_p^2=1/(16\pi G)$ and $\xi_d = (n-2)/4(n-1)$ which in four dimensions
$d=4$ is equal to 0 for minimal coupling and 1/6 for conformal coupling.
Adding a surface term in the action proportional to $K$,
the trace of the extrinsic curvature, is necessary for a consistent
variational theory \cite{surface} in the treatment of the
backreaction problem.

In the spatially- flat Friedmann-Robertson-Walker (FRW) universe with metric
\begin{equation}
ds^2=a^2(\eta)\Bigl(d\eta^2-\sum_i dx^2_i\Bigr)
\end{equation}
$R=6\ddot{a}/a^3, K=3\dot{a}/a^2$ (where a dot denotes a derivative
with respect to conformal time $\eta = \int dt/a$) we have
\begin{equation}
S_g=-Vl_p^2\int d\eta~(6\dot{a}^2+2\Lambda a^4)
\end{equation}
\begin{equation}
S_m=\frac{l_p^2}{2}\int d^4x\left[(\dot{\chi})^2-\sum_{i}(\chi_{,i})^2
-2(1-6\xi)\frac{\dot{a}}{a}\chi\dot{\chi}
-\left(m^2 a^2+(6\xi -1)\frac{\dot{a}^2}{a^2}\right)\chi^2\right].
\end{equation}
Here $\chi=a\Phi$ is the rescaled field variable  and $V$ is the volume
of the universe. From now on we will absorb $l_p$ by rescaling $\chi$ and $a$.

We can expand the scalar field in a box of co-moving volume $V$
(fixed coordinate volume)
\begin{equation}
\chi(x)=\sqrt{\frac{2}{V}}\sum_{\vec{k}}[q_{\vec{k}}^+
 \cos\vec{k}\cdot\vec{x} + q_{\vec{k}}^- \sin\vec{k}\cdot\vec{x}]
\end{equation}
which leads to the Lagrangian
\begin{equation}
L (\eta)=\frac{1}{2}\sum_{\sigma}^{+-}\sum_{\vec{k}}
\left[(\dot{q}_{\vec{k}}^{\sigma})^2-2(1-6\xi_d)\frac{\dot{a}}{a}
q_{\vec{k}}^{\sigma}\dot{q}_{\vec{k}}^{\sigma}
-\left(k^2+m^2 a^2+(6\xi_d -1)\frac{\dot{a}^2}{a^2}
\right)q_{\vec{k}}^{\sigma2}\right]
\end{equation}
where $k=|\vec{k}|$ and $S_m(\eta)=\int L(\eta)d\eta$.
The canonical momentum is
\begin{equation}
p_{\vec{k}}^{\sigma}=\frac{\partial L (\eta)}{\partial\dot{q}_{\vec{k}}^
{\sigma}}=\dot{q}_{\vec{k}}^{\sigma}-(1-6\xi_d )\frac{\dot{a}}{a}q_{\vec{k}}
^{\sigma}.
\end{equation}
Defining the canonical Hamiltonian the usual way we find
\begin{equation}
H (\eta)=\frac{1}{2}\sum_{\sigma}^{+-}\sum_{\vec{k}}
\left[p_{s\vec{k}}^{\sigma 2}+(1-6\xi_d )\frac{\dot{a}}{a}(p_{s\vec{k}}
^{\sigma}q_{\vec{k}}^{\sigma}+q_{\vec{k}}^{\sigma}p_{s\vec{k}}^{\sigma})
+\left(k^2+m^2 a^2+6\xi_d (6\xi_d -1)\frac{\dot{a}^2}{a^2}\right)q_{\vec{k}}
^{\sigma 2}\right]
\end{equation}
where the sum is over positive $k$ only since we have an expansion
over standing rather than travelling waves.
The system is quantized by promoting $(p_{\vec{k}}^{\sigma},q_{\vec{k}}
^{\sigma})$
to operators obeying the usual harmonic oscillator commutation relations.
In this way the dynamics of the field is reduced to the dynamics of
time-dependent harmonic oscillators. (See \cite{HM2} for details.)

In the case of a free quantized scalar field coupled to a spatially-
flat FRW universe with scale factor $a(s)$ the action thus belongs to the
general form
\begin{equation}
S[a,{\bf q}]
 = \int\limits_{t_i}^tds \Biggl[ L(a,\dot{a},s)
+  \sum_k\Bigl\{  {1\over 2}m_k(a,\dot{a})
  \Bigl(\dot q_k^2+b_k(a,\dot{a})q_k\dot{q}_n-
\omega^2_k(a,\dot{a})q_k^2\Bigr) \Bigr\}\Biggr].
\end{equation}
By tracing out the scalar field we can obtain an influence functional
and from which derive an equation of motion
for the scale factor in the semiclassical regime. Here since
we work explicitly in the semiclassical regime,
the environment is quantum and gravity enters classically
through the scale factor $a$.

We want to calculate the influence functional for this model. From (2.9) we
see that it is formally given by
\begin{equation}
{\cal F}[a,a']=\prod_k Tr\Bigl(\hat{U}_k[a_{t,t_i}]
\hat{\rho}_b (t_i)\hat{U}_k^{\dag}[a'_{t,t_i}]\Bigr)
\end{equation}
where $\hat{U}_k$ is the quantum propagator for the bath mode in (4.10) with
$a(s)$
treated as an arbitrary classical time dependent function. We have derived
this propagator before \cite{Mat} and will only quote the results here.
The result for a particular mode is
(we drop the mode label)
\begin{equation}
\hat{U}[a_{t,t_i}]=\hat{S}(r,\phi)\hat{R}(\theta)
\end{equation}
where
\begin{equation}
\hat{R}(\theta)=e^{-i\theta \hat{B}},\;\;\;\;\;
\hat{S}(r,\phi)=\exp [r(\hat{A}e^{-2i\phi}-\hat{A}^{\dag}e^{2i\phi})]
\end{equation}
and
\begin{equation}
\hat{A}=\frac{\hat{a}^2}{2},\;\;\;\;\;\hat{A}^{\dag}=
\frac{\hat{a}^{\dag 2}}{2},\;\;\;\;\;\hat{B}=\hat{a}^{\dag}\hat{a}+1/2.
\end{equation}
$\hat{S}$ and $\hat{R}$ are called squeeze and rotation operators
respectively.
The parameters $r,\phi,\theta$ are determined from the equations
\begin{eqnarray}
\dot{\alpha} & = & -ig^*\beta-ih\alpha \\
\dot{\beta} & = & ih\beta+ig\alpha
\end{eqnarray}
where
\begin{equation}
\alpha=e^{-i\theta}\cosh r,\;\;\;\;\;\beta=-e^{-i(2\phi+\theta)}\sinh r
\end{equation}
and
\begin{equation}
g(t)=\frac{1}{2}\left(\frac{m(t)\omega^2(t)}{c}
+\frac{m(t)b^2(t)}{4c}-\frac{c}{m(t)}+ib(t)\right)
\end{equation}
\begin{equation}
h(t)=\frac{1}{2}\left(\frac{c}{m(t)}+\frac{m(t)\omega^2(t)}{c}
+\frac{m(t)b^2(t)}{4c}\right).
\end{equation}
$c$ is an arbitrary positive real constant that is usually chosen so that
$g=0$ at $t_i$. We must have $\alpha=1$ and $\beta=0$ at $t_i$ so that the
initial condition for the propagator is satisfied. The time dependence on
$g$ and $h$ comes directly from $a$ in (4.10).

Applying (4.12) to (4.11) we find that the influence functional for a mode in
an initial vacuum state is given by
\begin{equation}
{\cal F}_k[a,a']=\sum_n\langle n|\hat{S}(r,\phi)\hat{R}(\theta)
|0\rangle\langle 0|\hat{R}^{\dag}(\theta')\hat{S}^{\dag}(r',\phi')|n\rangle
\end{equation}
where $|n\rangle$ are the usual number states. Using
\begin{equation}
\hat{R}(\theta)|0\rangle=e^{-i\theta/2}|0\rangle
\end{equation}
we find
\begin{equation}
{\cal F}_k[a,a']=\sum_n\Bigl[\langle n|\hat{S}(r,\phi)
|0\rangle\langle 0|\hat{S}^{\dag}(r',\phi')|n\rangle\Bigr]
e^{-i(\theta-\theta')/2}.
\end{equation}
With
\begin{equation}
\hat{S}(r,\phi)|0\rangle=(\cosh r)^{-1/2}\sum_{n=0}^{\infty}\left[
(-e^{2i\phi}\tanh r)^n\frac{\sqrt{(2n)!}}{2^n n!}|2n\rangle\right]
\end{equation}
and making use of
\begin{equation}
\sum_n\left[\frac{(2n)!}{(n!)^2}x^n\right]=\frac{1}{\sqrt{1-4x}}
\end{equation}
we can show that
\begin{equation}
{\cal F}[a,a']=\prod_k\frac{1}{\sqrt{\alpha_k
[a']\alpha_k^*[a]-\beta_k[a']\beta_k^*[a]}}.
\end{equation}
This shows yet another way of deriving the influence functional in terms of the
Bogolubov coeffients, in addition to the derivations  given in \cite{nfsg}.

\section{Einstein-Langevin Equation}

{}From the Hamiltonian (4.9) we see that the system- environment interaction is
separable for two cases: the massive conformally coupled field (for which
$\sigma=a^2$ in (3.1)) and the massless minimally coupled field
($\sigma=\dot{a}/a$) which also describes gravitons. For these
two cases the results from Sec. 3 apply: (3.12) is the appropriate
equation describing backreaction of the quantum scalar field on the metric.
To derive the Einstein-Langevin equation we need to compute the first
two cumulants given by (3.3) using the influence functional (4.25).

The solution of (4.15) and (4.16) can be written as
\begin{equation}
{\bf U}[a_{t,t_i}]={\cal T}\exp\left(-i\int_{t_i}^tds~{\bf u}(s)\right)
\end{equation}
where
\begin{equation}
{\bf u}(s)=\left(\begin{array}{cc}
h(s) & g^*(s) \\
-g(s) & -h(s)
\end{array} \right)
\end{equation}
and
\begin{equation}
{\bf U}[a_{t,t_i}]= \left(\begin{array}{cc}
\alpha [a_{t,t_i}] & \beta^* [a_{t,t_i}] \\
\beta [a_{t,t_i}] & \alpha^* [a_{t,t_i}] \end{array} \right).
\end{equation}
The key to calculating the functional derivative of (5.1) is recognizing that
we can always write
${\bf U}[a_{t,t_i}]={\bf U}[a_{t,\tau}]{\bf U}[a_{\tau,t_i}]$.
We therefore find
\begin{equation}
\frac{\delta {\bf U}[a_{t,t_i}]}{\delta \Delta(\tau)}=
\frac{\delta {\bf U}[a_{t,\tau}]}{\delta \Delta(\tau)}{\bf U}[a_{\tau,t_i}]+
{\bf U}[a_{t,\tau}]\frac{\delta {\bf U}[a_{\tau,t_i}]}{\delta \Delta(\tau)}.
\end{equation}
Making use of the formal expression for the time ordered representation of
(5.1) it is easy to see that
\begin{equation}
\frac{\delta {\bf U}[a_{t,\tau}]}{\delta \Delta(\tau)}=-i{\bf U}[a_{t,\tau}]
\int_{\tau}^tds ~\frac{\delta {\bf u}(s)}{\delta \Delta(\tau)}
\end{equation}
\begin{equation}
\frac{\delta {\bf U}[a_{\tau,t_i}]}{\delta \Delta(\tau)}=
-i\left(\int_{t_i}^{\tau}ds ~\frac{\delta {\bf u}(s)}{\delta \Delta(\tau)}
\right){\bf U}[a_{\tau,t_i}].
\end{equation}
Substituting (5.5) and (5.6) into (5.4) we find that
\begin{equation}
\frac{\delta {\bf U}[a_{t,t_i}]}{\delta \Delta(\tau)}=
-i{\bf U}[a_{t,\tau}]\left(\int_{t_i}^{t}ds
{}~\frac{\delta {\bf u}(s)}{\delta \Delta(\tau)}\right){\bf U}[a_{\tau,t_i}].
\end{equation}

\subsection{Massive conformally coupled field}
For the massive conformally coupled case we have $\sigma = a^2$ and
\begin{equation}
g=\frac{1}{2}\left[\frac{(k^2+m^2a^2)l_p^2}{c}-\frac{c}{l_p^2}\right],\;\;\;
h=\frac{1}{2}\left[\frac{(k^2+m^2a^2)l_p^2}{c}+\frac{c}{l_p^2}\right]
\end{equation}
in (4.18-19). From (3.3) and (4.25) (we have reinstated the Planck length)
we find the first cumulant of the stochastic force is
\begin{equation}
C_1(\eta;a_{\e,\e_i}^2]=-\frac{l_p^2m^2}{2}\sum_{\sigma}^{+-}
\sum_{\vec{k}}\langle \hat{q}_\e^2\rangle=
-\frac{l_p^2m^2\hbar}{4}\sum_{\sigma}^{+-}
\sum_{\vec{k}}\frac{1}{c}(\a_{\e}+\b_{\e})(\a_{\e}+\b_{\e})^*
\end{equation}
where $\hat{q}_{\e}^2=\hat{U}^{\dag}[a_{\e,\e_i}]\hat{q}^2\hat{U}[a_{\e,\e_i}]
$ and the average is with respect to
the vacuum. The propagator $\hat{U}$ is given by (4.12) with the Bogolubov
coefficients determined via (4.15-16) with $g,h$ given by (5.8).
We will use this notation below as well.
Similarly for the second cumulant we find
\begin{eqnarray}
C_2(\e,\e';a_{\e,\e_i}^2,a_{\e',\e_i}^2] & = &
-\frac{l_p^4m^4}{8}\sum_{\sigma}^{+-}
\sum_{\vec{k}}\Bigl[\langle \hat{q}_{\e}^2 \hat{q}_{\e'}^2\rangle+\langle
\hat{q}_{\e'}^2 \hat{q}_{\e}^2\rangle
-2\langle \hat{q}_{\e}^2\rangle\langle \hat{q}_{\e'}^2\rangle\Bigr]
\nonumber \\
& = & -\frac{l_p^4\hbar^2 m^4}{16}\sum_{\sigma}^{+-}
\sum_{\vec{k}}\frac{1}{c^2}\Bigl[(\b_{\e}+\a_{\e})^2(\a_{\e'}^*+\b_{\e'}^*)^2
\nonumber \\
&+&
(\b_{\e}^*+\a_{\e}^*)^2(\a_{\e'}+\b_{\e'})^2\Bigr].
\end{eqnarray}
Applying (3.16) to (5.9) we find the dissipation kernel to be
\begin{eqnarray}
\mu(\e,\e') & = & \frac{il_p^4 m^4}{4\hbar}\sum_{\sigma}^{+-}
\sum_{\vec{k}}\Bigl[\langle q_{\e}^2 q_{\e'}^2\rangle-\langle q_{\e'}^2
q_{\e}^2\rangle
\Bigr]\bigg|_{a^2=a_b^2} \nonumber \\
& = & \frac{il_p^4\hbar m^4}{8}\sum_{\sigma}^{+-}
\sum_{\vec{k}}\frac{1}{c^2}\Bigl[(\b_{\e}+\a_{\e})^2(\a_{\e'}^*+\b_{\e'}^*)^2-
(\b_{\e}^*+\a_{\e}^*)^2(\a_{\e'}+\b_{\e'})^2\Bigr]\bigg|_{a^2=a_b^2}.
\end{eqnarray}
Again we see the close relation between the noise and dissipation kernels.

Using (4.4) and $\sigma=a^2$ we find that the equation of motion (3.12) with
the background approximation becomes
\begin{equation}
\ddot{a}-\frac{2}{3}\Lambda a^3+\frac{a(\e)}{6Vl_p^2}
\Bigl[C_1(\e;a^2_{\e,\e_i}]|_{a^2=a^2_b}+
\int_{\e_i}^{\e} d\e' \tilde{a}^2(\e')\mu(\e,\e')\Bigr]=
-\frac{a(\e)}{6Vl_p^2}\xi(\e)
\end{equation}
where $\xi$ is a zero mean gaussian stochastic force with the correlator (5.10)
evaluated on the background $a_b$.

\subsection{Massless minimally coupled case}
For the massless minimally coupled case, $\sigma=\dot{a}/a$,

\begin{equation}
g=-i\frac{\dot{a}}{a}+\frac{1}{2}\left(\frac{l_p^2k^2}{c}-
\frac{c}{l_p^2}\right),
\;\;\;h=\frac{1}{2}\left(\frac{l_p^2 k^2}{c}+\frac{c}{l_p^2}\right),
\end{equation}
we get
\begin{equation}
C_1(\e;(\dot{a}/a)_{\e,\e_i}]=-\frac{1}{2}\sum_{\sigma}^{+-}\sum_{\vec{k}}
\langle (pq+qp)_{\e}\rangle
=-\frac{i\hbar}{2}\sum_{\sigma}^{+-}\sum_{\vec{k}}
[\a_{\e}^*\b_{\e}-\a_{\e}\b_{\e}^*]
\end{equation}
where $p$ is the canonical momentum from the Lagrangian (4.7) with $m=\xi=0$.
For this case $\frac{\partial \sigma}{\partial a}-
\frac{d}{d\e}\frac{\partial \sigma}{\partial\dot{a}}=0$ so we see from (3.12)
we must find $\dot{C}_1$. Taking the derivative of (5.14) and using (4.15-16)
and (5.13) (with $c=l_p^2k$) we find
\begin{equation}
\dot{C}_1(\e;(\dot{a}/a)_{\e,\e_i}]=\hbar \sum_{\sigma}^{+-}
\sum_{\vec{k}}k[\a_{\e}^*\b_{\e}+\a_{\e}\b_{\e}^*].
\end{equation}

For the second cumulant we find
\begin{eqnarray}
C_2(\e,\e';(\dot{a}/a)_{\e,\e_i},(\dot{a}/a)_{\e',\e_i}] &=&
\frac{1}{8}\sum_{\sigma}^{+-}
\sum_{\vec{k}}\Bigl[  \langle  (pq+qp)_{\e} (pq+qp)_{\e'}\rangle+
\langle(pq+qp)_{\e'} (pq+qp)_{\e}\rangle \nonumber \\
& - & 2\langle(pq+qp)_{\e}\rangle\langle (pq+qp)_{\e'}\rangle\Bigr]
\nonumber \\
& = & \frac{\hbar^2}{4}\sum_{\sigma}^{+-}
\sum_{\vec{k}}\Bigl[(\a_{\e}^2-\b_{\e}^2)(\a_{\e'}^{*2}-\b_{\e'}^{*2})
\nonumber \\
&+& (\a_{\e}^{*2}-\b_{\e}^{*2})(\a_{\e'}^2-\b_{\e'}^2)\Bigr].
\end{eqnarray}
{}From (3.17) and (5.15) the dissipation kernel is given by
\begin{equation}
\dot{\mu}(\e,\e') =  -\hbar\sum_{\sigma}^{+-}
\sum_{\vec{k}}k\Bigl[(\b_{\e}^2+\a_{\e}^2)(\a_{\e'}^{*2}-\b_{\e'}^{*2})+
(\b_{\e}^{*2}+\a_{\e}^{*2})(\a_{\e'}^2-\b_{\e'}^2)\Bigr]\bigg|_{\dot{a}/a=
(\dot{a}/a)_b}.
\end{equation}
The equation of motion (3.12) with the background approximation becomes
\begin{equation}
\ddot{a}-\frac{2}{3}\Lambda a^3-\frac{1}{12Vl_p^2a(\e)}\Bigl[\dot{C}_1(\e,
(\dot{a}/a)_{\e,\e_i})|_{\dot{a}/a=(\dot{a}/a)_b}+\int_{\e_i}^{\e}d\e'~
\frac{\dot{\tilde{a}}(\e')}{\tilde{a}(\e')}\dot{\mu}(\e,\e')\Bigr]=
\frac{\dot{\xi}(\e)}{12Vl_p^2a(\e)}.
\end{equation}

We need to know the stochastic properties of $\dot{\xi}(\e)$ given that
$\xi(\e)$ is a zero mean gaussian stochastic force with the correlator (5.16)
evaluated on a background.
We can deduce this by integrating by parts the noise term in the effective
action (3.4). We find that (relaxing the notation for $C_2$)
\begin{eqnarray}
\int_{\e_i}^{\e_f}d\e_1d\e_2\Delta(\e_1)\Delta(\e_2)
C_2(\e_1,\e_2]& = &~{\rm surface~term}~ \nonumber \\
&+& \int_{\e_i}^{\e_f}d\e_1\Gamma(\e_1)
\biggl[\frac{dC_2}{d\e_1}(\e_1,\e_i)\Gamma(\e_i)-
\frac{dC_2}{d\e_1}(\e_1,\e_f)\Gamma(\e_f) \nonumber \\
&+&\frac{dC_2}{d\e_1}(\e_i,\e_1)\Gamma(\e_i)-
\frac{dC_2}{d\e_1}(\e_f,\e_1)\Gamma(\e_f)\biggr] \nonumber \\
&+&\int_{\e_i}^{\e_f}dt_1\int_{\e_i}
^{\e_f}d\e_2\Gamma(\e_1)\Gamma(\e_2)
\frac{d^2C_2(\e_1,\e_2]}{d\eta_1 d\eta_2}
\end{eqnarray}
where $\Gamma(\e)=\int d\e~\Delta(\e)={\rm ln}~a-{\rm ln}~a'$.
The surface term will
not contribute to the equation of motion but the last term of (5.19) shows
clearly that
$\dot{\xi}(t)$ corresponds to a zero mean gaussian stochastic force with the
correlator
\begin{equation}
C_{2\dot{\xi}}(\e,\e']=\frac{d^2C_2(\e_1,\e_2]}{d\eta_1 d\eta_2}.
\end{equation}
The meaning of the middle term of (5.19)
is more difficult to interpret. It vanishes only when the noise
is stationary since we then have $C_2(\e,\e']=C_2(\e-\e']$.
We will not discuss this term further since it will vanish in the example we
consider next. Clearly though, its meaning will need to be considered for a
study about nonstationary backgrounds.

\subsection{Backreaction of graviton fluctuations about flat space}
A simple case to study is a massless minimally- coupled
field around a flat background ($\tilde{a}=a$). In this case $\alpha(\e)
=e^{-ik\e}$ and
$\beta(\e)=0$. We see that in this case the first cumulant (5.14) vanishes.
This should be compared to the massive field where the first cumulant is
divergent around a flat background.
The noise kernel (5.16) becomes
\begin{equation}
C_2(\e-\e']  = \frac{\hbar^2 V}{32\pi^2}\int_{0}^{\infty}dk
{}~k^2\cos[k(\e-\e')]=-\frac{\hbar^2 V}{32\pi}\delta ''(\e-\e')
\end{equation}
where a prime on a function denotes a derivative taken with respect to its
argument. From (5.20) we have
\begin{equation}
C_{2\dot{\xi}}(\e-\e'] = \frac{\hbar^2 V}{32\pi^2}\int_{0}^{\infty}dk
{}~k^4\cos[k(\e-\e')]=\frac{\hbar^2 V}{32\pi}\delta ''''(\e-\e').
\end{equation}
The dissipation kernel (5.17) becomes
\begin{equation}
\dot{\mu}(\e-\e') = -\frac{\hbar V}{16\pi^2}\int_{0}^{\infty}dk
{}~k^3\cos[k(\e-\e')].
\end{equation}

The Einstein-Langevin equation (5.18) becomes
\begin{equation}
\ddot{a}-\frac{2}{3}\Lambda a^3-\frac{1}{12Vl_p^2a(\e)}\int_{\e_i}^{\e}
d\e'~\frac{\dot{a}(\e')}{a(\e')}\dot{\mu}(\e-\e')=
\frac{1}{12Vl_p^2a(\e)}\dot{\xi}(\e).
\end{equation}
where $\dot{\xi}$ is a zero- mean Gaussian force with the correlator (5.22).
The solution of the Einstein-Langevin equations discussed here are beyond
the scope of the present paper. We plan to consider these solutions in the
future in the context of a general study into the dynamics of second order
Langevin equations with non-local dissipation and colored noise \cite{matacz}.
\section{Summary}

Together with two related work \cite{nfsg,fdrc}, this paper seeks
to establish a new framework for the study of semiclassical gravity theory
based on the Einstein-Langevin equation. In \cite{nfsg} the noise and
fluctuation terms are identified from the closed-time-path formalism
and the Einstein-Langevin equation derived for perturbances off the
Robertson-Walker spacetime.
In \cite{fdrc} the influence functional method is used to derive an
equation of motion for the anisotropy matrix of the Bianchi Type-I universe.
Dissipation of anisotropy from particle creation in a quantum scalar field
is seen to be driven by an additional stochastic source (noise) term
related to the fluctuations of particle creation and shown to be a
manifestation
of a fluctuation-dissipation relation. In this paper, we have derived the
following results:
\begin{itemize}
\item By carrying out a functional Taylor series expansion on the influence
functional we show
how the successive orders measure the higher cumulants of noise
in its most general (colored and multiplicative) forms, the lowest order
truncation yielding a Gaussian noise.
The second cumulant gives the autocorrelation function for the
stochastic force (noise), which drives the Einstein-Langevin equation.

\item Using a general form for the Hamiltonian of a quantum field whose normal
modes are coupled to a curved spacetime parametrically, we showed a new way
to derive the influence functional in terms of the Bogolubov coefficients
between the second-quantized operators of Fock spaces at two different times.
This relation  connects our new influence functional / effective action
method with the traditional canonical quantization approach and thus
incorporates the established body of knowledge in quantum field theory
in curved spacetimes.

\item With the previous two results we were able to express the noise and
dissipation
kernels in terms of the Bogolubov coefficients. This connection offers a more
transparent interpretation of the physical meaning of the many statistical
mechanical processes such as decoherence and dissipation in terms of
particle creation and related quantum effects.

\item We have also derived the form of the Einstein-Langevin equations
for some well-studied cases of scalar fields in Robertson-Walker and
de Sitter spacetimes. They form the starting points of the next stage
of work, which is  the solution of these equations for the analysis of
fluctuations, instability and phase transition.
We hope to report on these problems in future communications.
\end{itemize}

{\bf Acknowledgements} We thank Esteban Calzetta and Sukanya Sinha for
discussions. This work was done when AM visited the relativity theory
group of the University of Maryland.
Research is supported in part by the National Science Foundation
under grant 91-19726.

\end{document}